\documentclass[lettersize,journal]{IEEEtran}
\usepackage{amsmath,amsfonts}
\usepackage{algorithmic}
\usepackage{algorithm}
\usepackage{array}
\usepackage[caption=false,font=normalsize,labelfont=sf,textfont=sf]{subfig}
\usepackage{textcomp}
\usepackage{stfloats}
\usepackage{url}
\usepackage{verbatim}
\usepackage{graphicx}
\usepackage{cite}
\usepackage{color}
\newcommand{\llr}[1]{{\color{black}#1}}
\usepackage{multirow}

\begin{document}

\title{\llr {A Tampering Risk of Fiber-Based Frequency Synchronization Networks and Its Countermeasures}}

\author{Hongfei Dai, Yufeng Chen, Wenlin Li, Fangmin Wang, Guan Wang, Zhongwang Pang, and Bo Wang
\thanks{This work was supported in part by the National Natural Science Foundation of China under Grant 62171249, in part by the National Key Project of Research and Development under Grant 2021YFA1402102, and in part by Tsinghua Initiative Scientific Research Program. \emph{(Corresponding author: Bo Wang.)}}
\thanks{The authors are with State Key Laboratory of Precision Space-time Information Sensing Technology, Department of Precision Instrument, Tsinghua University, Beijing 100084, China, and also with the Key Laboratory of Photonic Control Technology (Tsinghua University), Ministry of Education, Beijing 100084, China (e-mail: bo.wang@tsinghua.edu.cn).}
}



\maketitle

\begin{abstract}
\llr { Fiber optic networks are used worldwide and have been regarded as excellent media for transmitting time-frequency (TF) signals. In the past decades, fiber-based TF synchronization techniques have been extensively studied. Instruments based on these techniques have been successfully applied. With the increasing application of TF synchronization instruments, their security has become an important issue. Unfortunately, the security risks of fiber-based frequency synchronization (FbFS) instruments have been overlooked. This paper proposes a frequency tampering method called “frequency lens”. On a 200 km fiber link, we demonstrate a frequency tampering scenario using a frequency lens-enabled frequency tampering module (FTM). On the user side, the frequency value of the recovered 100 MHz signal can be stealthily altered within a range of 100 MHz±100 Hz, while the frequency dissemination stability of the system remains normal. Related to this tampering risk, potential hazards in three different application scenarios, which rely on precise frequency references, are analyzed. Two countermeasures are also proposed to solve this tampering risk.}

\end{abstract}



\section{Introduction}\label{sec:introduction}
\begin{figure}[!h]
	\centering
	\includegraphics[width=3.5in]{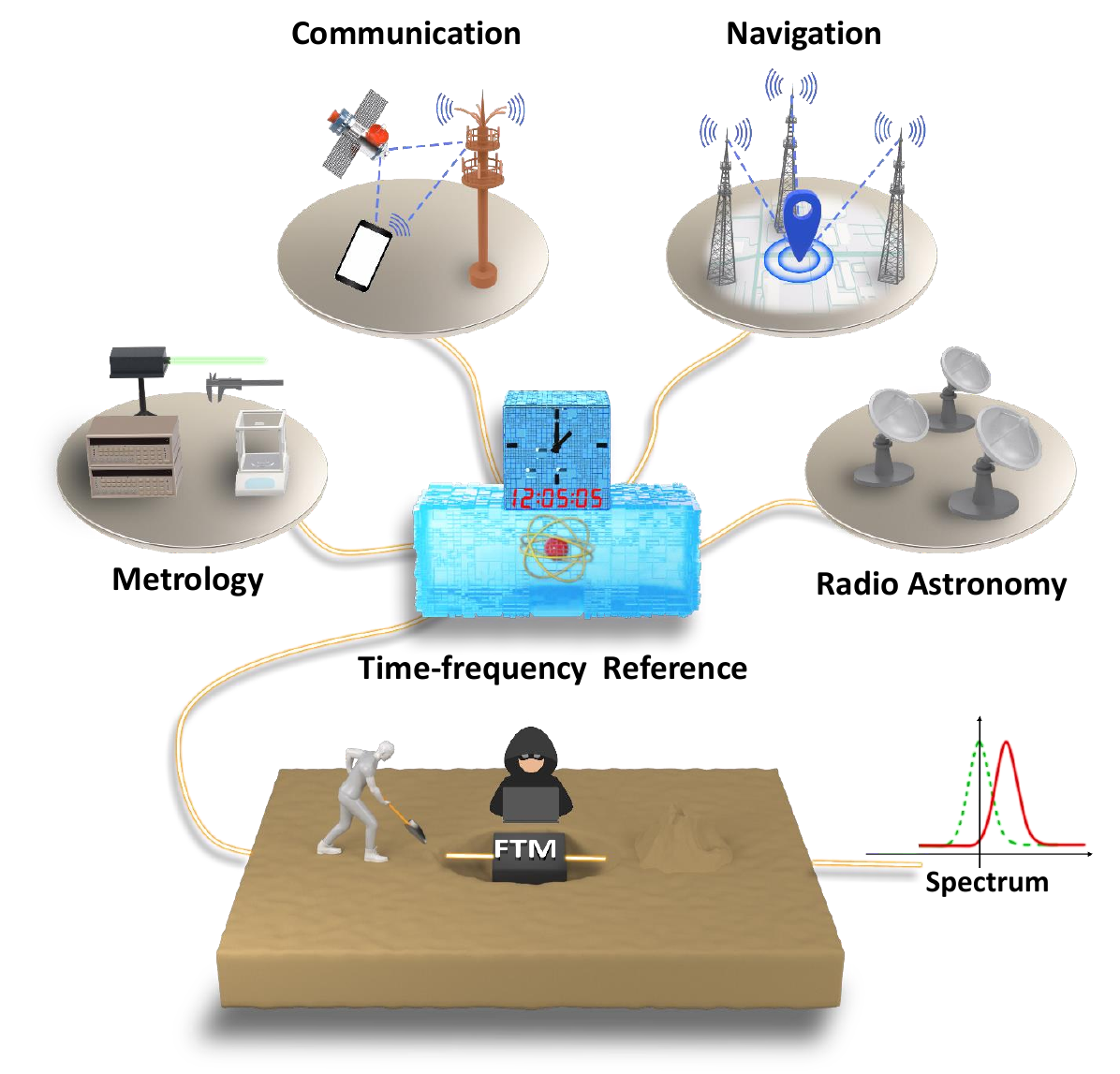}
	\caption{Schematic diagram of the fiber-based TF synchronization network and tampering risk behavior. FTM, frequency tampering module.}
	\label{fig1}
\end{figure}

\IEEEPARstart{A}t present, optical fiber networks are used worldwide and have been one of the largest infrastructures for human utilization.
\llr{As an excellent transmission medium, the optical fiber has unique advantages in transmitting time-frequency (TF) reference. In the past decades, fiber-based TF synchronization techniques have been widely studied \cite{Ye2003,Williams2008,Masaki2008,Wang2012,Bai2013,Lisdat2016,Rizzi2016,Riehle2017,Wang2020JL,Liu2022Bo}. The related instruments have been successfully applied in various fields, such as metrology \cite{Lisdat2016,Riehle2017,Schioppo2022}, communication \cite{johnson2020}, navigation \cite{Lewis2021}, and radio astronomy \cite{He2018,Chen2021}, as shown in Fig. \ref{fig1}. In the near future, fiber-based TF synchronization instruments will support strict timing requirements in the fields of fifth-generation/sixth-generation communication \cite{Lukasz2020}, smart cities \cite{Zhou2023}, global seismic monitoring \cite{Marra2018,Grotti2018}, and distributed computation \cite{Clark2020}.}

\llr{With the increasing applications of fiber-based TF synchronization instruments, their security has become an important issue that must be considered in advance \cite{narula2018}. There have been previous studies on attacks and countermeasures related to fiber-based time synchronization (FbTS) systems \cite{lee2019,Li2023,Xu2023,Zhang2021,Liu2022}. However, research on corresponding aspects of fiber-based frequency synchronization (FbFS) systems is limited. Notably, those methods designed for attacking FbTS cannot be directly applied to FbFS. The primary method in attacking FbTS is to partially disrupt the symmetry of the fiber link in both forward and backward directions, causing asymmetry in the delay of time signals within the link. However, the asymmetry of a small portion of the fiber link cannot cause a disastrous effect on frequency synchronization. For FbFS instruments, the disastrous attack is covertly and slightly tampering with the disseminated frequency reference without affecting its stability. This means that third parties can control important applications with a frequency tampering module (FTM), as shown in Fig. \ref{fig1}. It would not disrupt the normal operation of applications relying on frequency synchronization but could lead to incorrect or biased results.}

\llr{ For FbFS systems, there are normally two phase-locked loops (PLLs), as shown in Fig. \ref{fig2}(a). PLL1 is used to compensate for the phase fluctuation induced by fiber dissemination. PLL2 is used as a clean oscillator to optimize the phase noise spectrum of the recovered frequency signal \cite{Wang2012,Chen2022}. Monitoring the locking status of PLL1 normally serves as an effective method for system security monitoring. PLL1 compares the returned frequency signal from the user site with the local frequency reference, making it sensitive to any alterations in the transmitted signal. Typically, tampering with the transmitted frequency signal results in a significant change in the locking status of PLL1. Although this monitoring method seems simple and effective, it has inherent flaws. Once a tampering method can make the returned frequency signal unbiased and bypass the locking status monitoring, tampering activities will become undetectable.}

\llr{In this paper, we propose a frequency tampering method called “frequency lens” and give countermeasures against it. With a frequency lens-enabled FTM, it can controllably alter the value of the disseminated frequency at any point along the fiber link. This alteration does not compromise the relative stability of frequency dissemination and can bypass locking status monitoring. In our case, tampering with the recovered 100 MHz signal within ±100 Hz is demonstrated on a 200 km fiber link. The relative stability of the recovered biased frequency (compared with the reference at the server site) is maintained at the same level as the normal state, i.e., $2.2\times 10^{-14}/1$ s and $2.2\times 10^{-17}/10^5$ s. We analyze its damaging effects on timekeeping, metrology, and astronomical observation, respectively. To solve the security issue of FbFS systems, we propose two countermeasures against frequency tampering methods. We anticipate this article will provoke research interest in the security issue of FbFS instruments and provide instructive insights for future investigations.}

\section{Method}
\begin{figure}[h]
	\centering
	\includegraphics[width=3.375in]{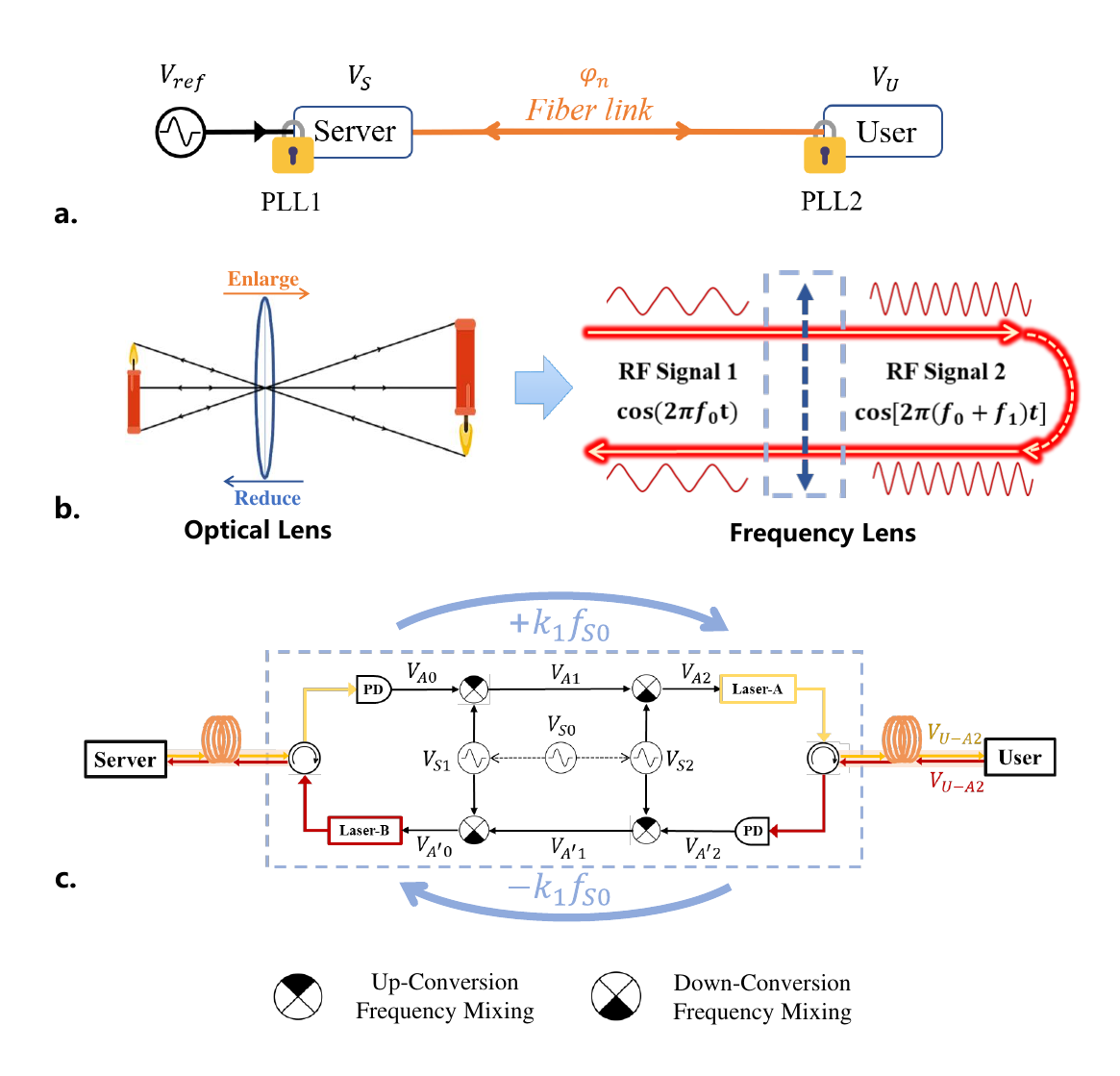}
	\caption{Principles of FbFS and the frequency lens. (a) Structure of a typical FbFS system. (b) Comparison between the optical lens and frequency lens. (c) Schematic diagram of the frequency lens. PD, photodetector.}
	\label{fig2}
    \vspace{-3mm}
\end{figure}

    We consider a typical fiber-based radio frequency synchronization system. The schematic diagram is shown in Fig. \ref{fig2}(a), which includes a reference $V_{ref}=\cos(2\pi  f_{ref} t+\phi_{ref} )$, a server $V_S=\cos(2\pi  f_S t+\phi_S )$, a user $V_U=\cos(2\pi  f_U t+ \phi_U ) $, and a fiber link with one-way loop noise $\phi_n$ \cite{Wang2012}. For convenience, the amplitude items of these signals are disregarded. When the frequency synchronization system is active by PLL1, the frequency $f_S$ is aligned with $f_{ref}$. The phase $\phi_S$ is locked to the difference between $\phi_{ref}$  and $\phi_n$, i.e., $\phi_S=\phi_{ref}-\phi_n$. After transmission, the signal $V_U$ is recovered at the user site, and the one-way loop noise is automatically eliminated. As a consequence, the following relationship can be applied:
\begin{equation}
	\phi_U = \phi_S + \phi_n = \phi_{ref},\label{eq1}
\end{equation} 
\begin{equation}
	f_U = f_S  =f_{ref}.\qquad\ \label{eq2}
\end{equation}

Under normal conditions, it enables highly stable frequency synchronization between the user and reference. If an intruder attempts to break the synchronization scenario, such as tampering with the recovered frequency, the locking status of PLL1 is affected. However, if the tampering operation can bypass the locking status monitoring, the recovered frequency can be slightly and covertly altered.

Here, we use a “frequency lens” to achieve this covert intrusion. The concept of the frequency lens is inspired by an optical lens. In geometrical optics, the reversibility principle states that light will follow exactly the same path if its direction of travel is reversed. An optical lens can form an enlarged image, as shown in Fig. \ref{fig2}(b). It can also form a reduced image by switching the position of the object and image. Thus, if the disseminated frequency can be changed in a similar way, the monitoring of the locking status will be bypassed, and stealthy frequency tampering will be achieved. Fig. \ref{fig2}(c) illustrates the detailed structure of the frequency lens. There is an oscillator $V_{S0}$, i.e., $V_{S0}=\cos(2\pi f_{S0} t+\phi_{S0} )$. The signals $V_{S1}$ and $V_{S2}$ are synthesized via $V_{S0}$, as follows:

\begin{equation}
\begin{aligned}
    	V_{S1} &= \cos( 2\pi f_{S1} t+\phi_{S1} )  \\
		& = \cos[ 2\pi (k+k_1 ) f_{S0} t+(k+k_1 ) \phi_{S0} ], \label{eq3}
\end{aligned}
\end{equation}
\begin{equation}
\begin{aligned}
	V_{S2} &= \cos( 2\pi f_{S2} t+\phi_{S2} ) \\
			& = \cos( 2\pi kf_{S0} t+k\phi_{S0} ), \qquad\qquad\quad\ \ \label{eq4}
      \end{aligned}
\end{equation}
where $k$ and $k_1$ are the frequency synthesis factors. Compared with $k$, the factor $k_1$ is very small, which is usually smaller than  $2.1\times 10^{-4}$.

Using the frequency lens, the input signal $V_{A0}$ (the disseminated RF signal modulated on the laser carrier, i.e., $V_{A0}=\cos(2\pi f_{A0} t+\phi_{A0})$) can be parametrically converted to the output signal $V_{A2}$ in the forward direction. Conversely, in the backward direction, the signal $V_{A^\prime 2}$ (the same frequency as $V_{A2}$) can be parametrically converted to the signal $V_{A^\prime 0}$ again (the same frequency as $V_{A0}$). Specifically, the parametrical conversion process is accomplished by using second-order nonlinear electronic devices, such as electric mixers, to generate signals $V_{A1}$ and $V_{A2}$. In the forward direction, $V_{A1}$ is generated by up-mixing $V_{A0}$ and $V_{S1}$, while $V_{A2}$ is generated by down-mixing $V_{A1}$ and $V_{S2}$, as follows:
\begin{IEEEeqnarray}{rCl}
    V_{A1} & =& \cos(2\pi f_{A1} t+\phi_{A1} ) \nonumber \\
	& =& \cos \left\{ 2\pi [(k+k_1 ) f_{S0}+f_{A0} ]t+(k+k_1 ) \phi_{S0}+\phi_{A0} \right\},\nonumber\\
    \label{eq5} \\
    V_{A2} & =& \cos(2\pi f_{A2} t+\phi_{A2} ) \nonumber \\
	& =& \cos[2\pi (k_1 f_{S0}+f_{A0} )t+k_1 \phi_{S0}+\phi_{A0} ].\label{eq6}
\end{IEEEeqnarray}
Consequently, the disseminated frequency is shifted from $f_{A0}$ to ($k_1 f_{S0}+f_{A0}$). The output signal $V_{A2}$ can be used to modulate the laser carrier again, which is transmitted to the user site along the optical fiber link. Referring to Appendix A, the RF signal $V_{U-A2}$ is recovered at the user side, i.e., $V_{U-A2}=\cos[2\pi (k_1 f_{S0}+f_{A0} )t+k_1 \phi_{S0}+\phi_{A0}+\phi_{n2} ]$, where $\phi_{n2}$ is the phase noise introduced by this part of the fiber link between the frequency lens and the user site. Notably, the application of the frequency lens induces $k_1 \phi_{S0}$ in phase noise. For example, if the frequency stability of $V_{SO}$ is $1\times 10^{-11}/1 s$ is chosen, $k_1 \phi_{S0}$ is $2\times 10^{-15}/1 s$ when $k_1$ is set as $2\times 10^{-4}$. It proves that it does not significantly affect the relative stability of the frequency dissemination.

At the user site, the RF signal $V_{U-A2}$ is also modulated on another laser carrier, whose wavelength is different, and returned to the frequency lens along the fiber. In the backward direction, signal $V_{A^\prime 2}$ is detected by the frequency lens, as follows:
\begin{IEEEeqnarray}{rCl}
V_{A^\prime 2} & =& \cos(2\pi f_{A^\prime 2} t+\phi_{A^\prime 2} ) \nonumber\\
	&=& \cos[2\pi (k_1 f_{S0}+f_{A0} )t \ +k_1 \phi_{S0}+\phi_{A0}+2\phi_{n2} ].\label{eq7} \qquad
\end{IEEEeqnarray}
In the lower part of Fig. \ref{fig2}(c), $V_{A^\prime 0}$ can be obtained by similar treatments to that of Eq. \eqref{eq5} and Eq. \eqref{eq6}, as follows:

\begin{equation}   
\begin{aligned}
	V_{A^\prime 1} & =\cos(2\pi f_{A^\prime 1} t+\phi_{A^\prime 1} )\\
	& =\cos \Bigl\{ 2\pi  \left[ (k+k_1 ) f_{S0}+f_{A0} \right]t\\
	&\qquad\quad +(k+k_1 ) \phi_{S0}+\phi_{A0}+2\phi_{n2} \Bigr\},\label{eq8}
\end{aligned}
\end{equation}
\begin{equation}    
\begin{aligned}
	V_{A^\prime 0} & =\cos(2\pi f_{A^\prime 0} t+\phi_{A^\prime 0} ) \\
	& =\cos( 2\pi f_{A0} t+\phi_{A0}+2\phi_{n2} ).\qquad\quad\label{eq9}
\end{aligned}
\end{equation}
In this way, we achieve symmetrical enlargement and reduction of the disseminated frequency, such as $f_{A0}\rightarrow f_{A2}$ and $f_{A^\prime 2}\rightarrow f_{A^\prime 0}$. Considering the relationship between $f_{A0}=f_{A^\prime 0}$ and $f_{A2}=f_{A^\prime 2}$, the insertion of the frequency lens does not affect the normal operation of the frequency synchronization system. In other words, the disseminated frequency can be altered by slightly changing the parameter $k_1$, and the locking status will be maintained.

The premise of this tampering operation is to covertly insert a tampering module with a frequency lens function into the fiber link. Considering the operation status of optical fiber networks, link outages due to natural disasters, human accidents, and extreme weather are common. Moreover, according to the above description, the intrusion position of the tampering module can be arbitrary along the fiber link. All of these conditions provide ample opportunities for tampering module intrusion, making the possibility of these risky situations much more likely.

\section{Experimental Setup}

\begin{figure}[h]
    \centering
    \includegraphics[width=3.375in]{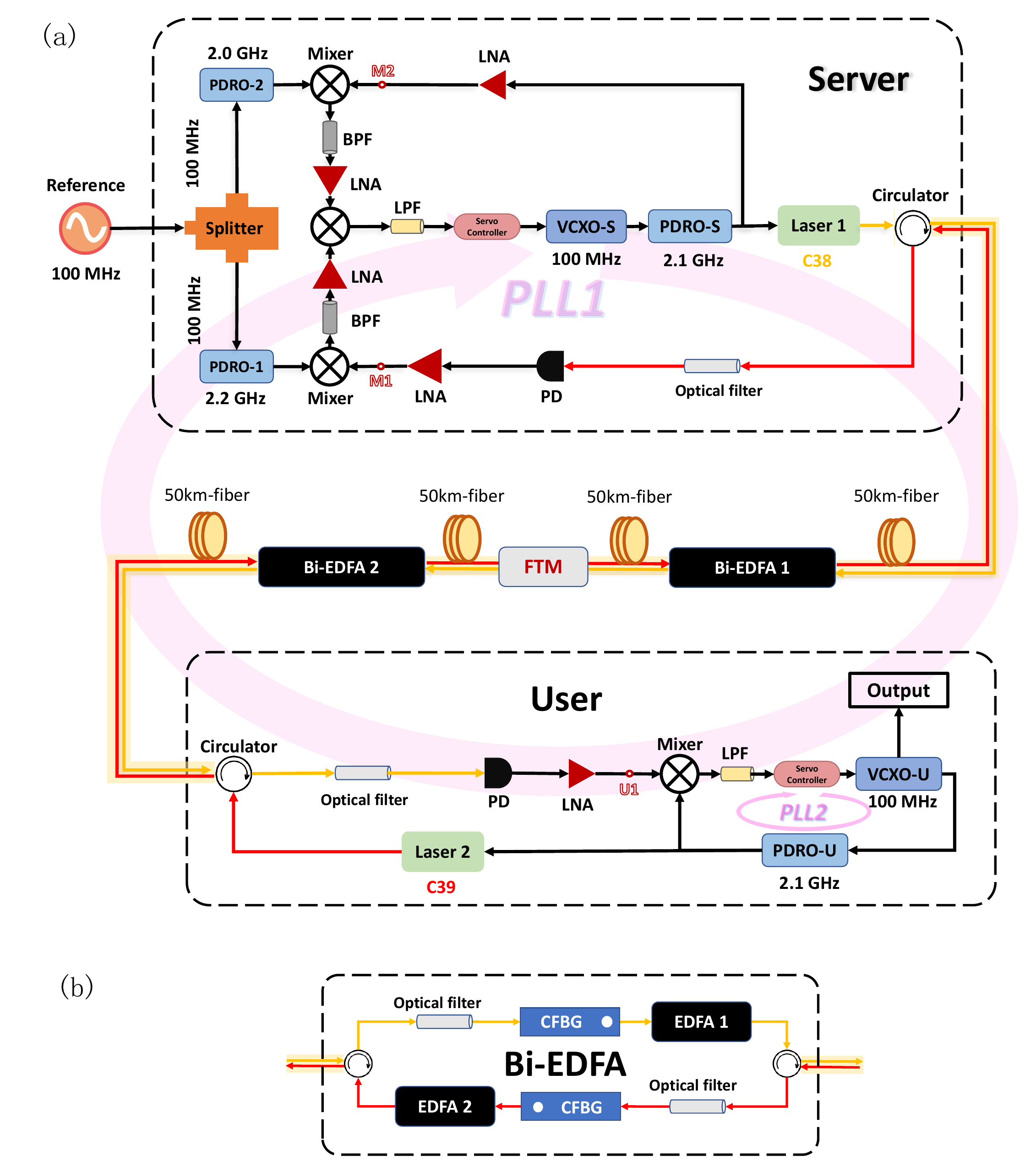}
    \caption{ \llr{Schematic diagram of the frequency synchronization system, bidirectional erbium-doped fiber amplifier (Bi-EDFA). (a) Frequency synchronization system, containing a server, a user and a fiber link. The black line represents RF signals. The yellow line represents the optical signal with the wavelength of 1546.92 nm (C38), and the red line represents the optical signal with the wavelength of 1546.12 nm (C39). (b) Chirped fiber grating enhanced Bi-EDFA, which plays the role of optical power amplification and dispersion elimination in the fiber link. EDFA, erbium-doped fiber amplifier; VCXO, voltage-controlled crystal oscillator; PDRO, phase-locked dielectric resonant oscillator; LNA, low noise amplifier; CFBG, chirped fiber grating.}}
\label{fig3}
\end{figure}

\llr{ As shown in Fig. \ref{fig3}(a), we have established a fiber-based radio frequency synchronization system. The frequency synchronization system consists of a server and a user, which are connected by a 200 km fiber link. The server is referenced to a 100 MHz output of a hydrogen maser. The system builds 2 PLLs to recovered a frequency signal at the user that is synchronized with the hydrogen maser. To improve the detection resolution of phase noise during transmission, the transmitted signal is 2.1 GHz RF signal modulated on the laser carriers, whose wavelengths are 1546.92 nm (C38) and 1546.12 nm (C39). A detailed description of the principles of this system can be found in Appendix A. For attenuation and dispersion, the erbium-doped fiber amplifier (EDFA) and the chirped fiber Bragg grating (CFBG) have emerged as effective solutions\cite{Chen2022}. In the experiment, we insert two CFBG-enhanced bidirectional EDFA (Bi-EDFAs) into the fiber link to amplify the laser signals transmitted in both directions and to compensate for dispersion. The structure of Bi-EDFA is shown in Fig. \ref{fig3}(b), and two CFBGs are used together with EDFA-1 and EDFA-2, respectively.}

\begin{figure*}[!t]
    \centering
	\includegraphics[width=6.75in]{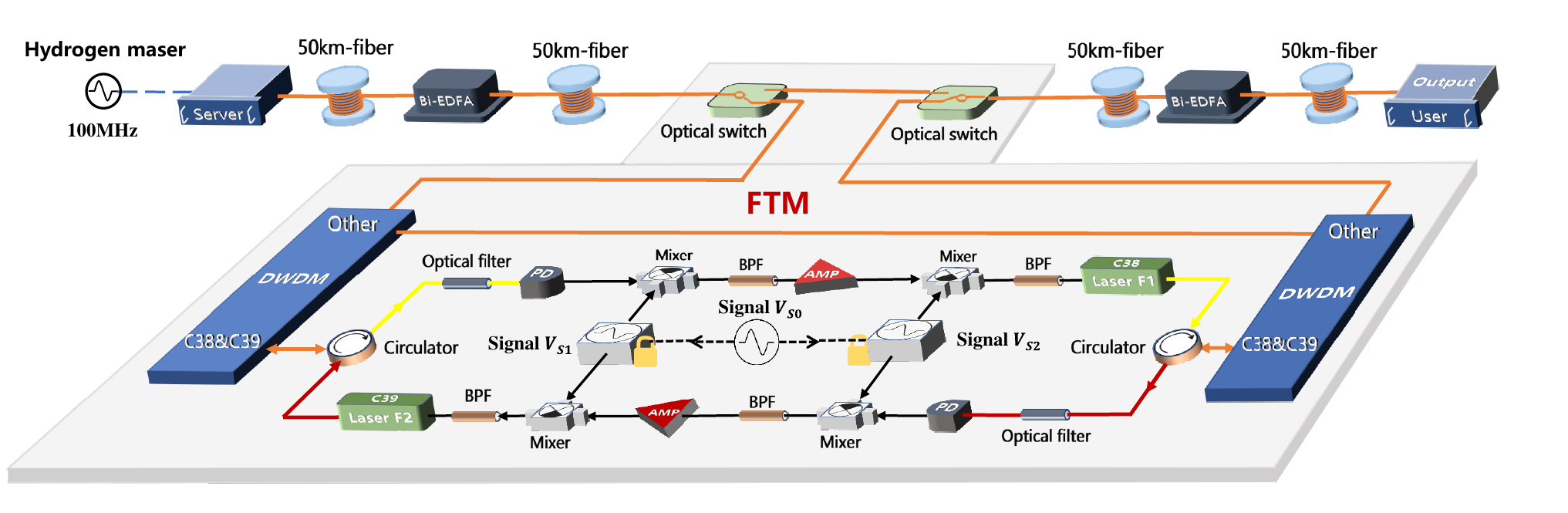}
	\caption{Schematic diagram of frequency lens-enabled FTM. The orange lines represent optical fiber links. The black line represents RF signals. DWDM, dense wavelength division multiplexing; PD, photodetector; BPF, band-pass filter; AMP, amplifier.}
	\label{fig4}
\end{figure*}

To realize the frequency tampering, we insert the frequency lens-enabled FTM into the fiber link, whose structure is shown in the gray board of Fig. \ref{fig4}. This FTM is composed of two dense wavelength division multiplexing (DWDM) modules, two optical switches, a short fiber connected to the ‘other’ ports of DWDM modules, and the key part of the FTM connected to the C38/C39 ports of DWDM modules. Inside the frequency lens, the optical signals from both directions are detected by photodetectors (PDs). A 10 MHz oscillator $V_{S0}$ is used to synthesize the signals $V_{S1}$ and $V_{S2}$ as shown in Eq. \eqref{eq3} and Eq. \eqref{eq4}, where $k=90$, and $k_1$ can be set within ±$2.1 \times 10^{-4}$. Specifically, $f_{S1}$ is set to $(900+10k_1)$ MHz, and $f_{S2}$ is set to 900 MHz.

\section{Result}
In our experiment, $k_1$ is the key to achieving frequency tampering, and $k$ is also selected to avoid harmonic interference in the frequency mixing process. Based on our experimental system, we obtain that the frequency $f_U$ recovered by the user satisfies the following equation:
\begin{equation}    
	f_U= \left(100+ \frac{10k_1}{21} \right) \textnormal{MHz},\label{eq10}
\end{equation}
whose derivation process is provided in Appendix A. For example, when $k_1$ is set as $2.1 \times 10^{-4}$, the recovered frequency at the user will be altered to 100 MHz+100 Hz. An experimental frequency tampering result is found in the video in Supplemental Material. We can observe from the video that the frequency value of the recovered 100 MHz signal can be stealthily altered within a range of 100 MHz±100 Hz over the 200 km fiber link. At the same time, the locking status remains normal. It is worth noting that, for numerous scientific applications, even small frequency offsets, as small as 1 Hz, are unacceptable. The 100 Hz offset is deliberately chosen in the experiment to illustrate the destructive capability of this method.

\begin{figure}[!h]
	\centering
	\includegraphics[width=2.8in]{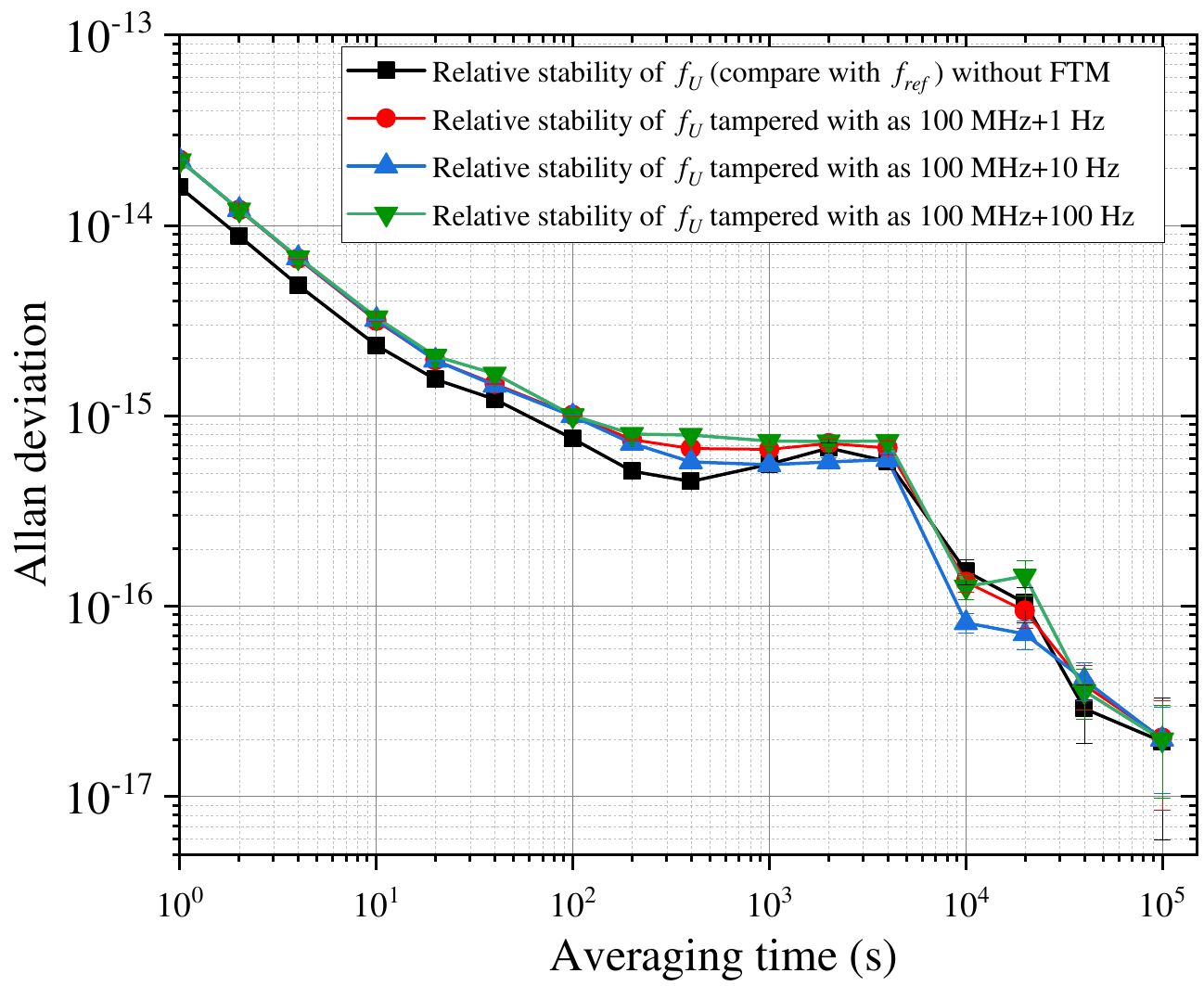}
	\caption{Relative stability of the recovered frequency signal under the different parameter settings of the FTM. With the FTM, $f_U$ is altered to 100 MHz + 1 Hz, 100 MHz + 10 Hz, and 100 MHz + 100 Hz, respectively. A comparison is made with the scenario without FTM insertion.}
	\label{fig5}
\end{figure}

Next, we further investigate the effect of the FTM on frequency dissemination stability. We select three different values of $k_1$ to set the corresponding recovered frequency $f_U$ to 100 MHz+1 Hz, 100 MHz+10 Hz, and 100 MHz+100 Hz. As a comparison, under the same experimental conditions, we also measure the frequency dissemination stability without the FTM insertion. The experimental results are shown in Fig. \ref{fig5}, and the duration of each test is approximately six days. When the FTM is not added to the fiber link, relative stability results are $1.4\times10^{-14}/1$ s and $2.0 \times 10^{-17}/10^5$ s. When the FTM is added, the relative stability results under three different values of $k_1$ remain at $2.2 \times 10^{-14}/1$ s and $2.2 \times 10^{-17}/10^5$ s. Thus, the insertion of the FTM has almost no effect on the frequency dissemination stability and the relative stability of the recovery signal is also almost the same when $k_1$ takes different values.

\begin{figure}[h]
	\centering
	\includegraphics[width=3.3in]{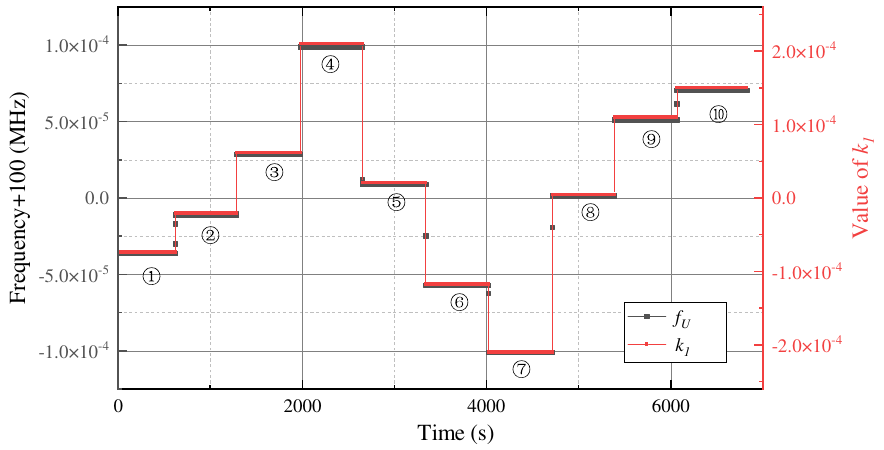}
	\caption{Variation in $f_U$ during the fast adjustment of $k_1$. $k_1$ is randomly changed 10 times, indicated by the black line.}
	\label{fig6}
\end{figure}

To show the characteristics of arbitrary adjustment and fast response of the frequency lens-enabled FTM, we carry out another frequency tampering experiment. The corresponding result is shown in Fig. \ref{fig6}, where the parameter $k_1$ is changed 10 times within 7000 s. The recovered frequency $f_U$ can be 

\noindent accurately and quickly altered, and the locking status of FbFS system is effectively maintained. Moreover, we measure the short-term relative stability of the recovered $f_U$ during each change, as shown in Table \ref{tab1}; thus, the relative stability of the recovered frequency is kept at the same level after the adjustment.

\vspace{-5mm}
\renewcommand\arraystretch{1.2}
\begin{table}[!h]
\caption{Short-term relative stability of the recovered frequency signal.}\label{tab1}
\centering
 \setlength\tabcolsep{2mm}{
\begin{tabular}{c|l|ccccc}
    \hline
    \multicolumn{2}{c|}{ Test sequence } & 1 & 2 & 3 & 4 & 5 \\
   \hline
   \multirow{2}{*}{ Allan Deviation } & $\times 10^{-14} / 1 \mathrm{~s}$ & 2.5 & 2.9 & 2.7 & 2.8 & 3.1 \\
	& $\times 10^{-15} / 100 \mathrm{~s}$ & 1.8 & 2.0 & 1.4 & 3.0 & 0.8\\
     \hline
    \multicolumn{2}{c|}{ Test sequence } & 6 & 7 & 8 & 9 & 10\\
   \hline
   \multirow{2}{*}{ Allan Deviation } & $\times 10^{-14} / 1 \mathrm{~s}$ & 3.4 & 3.3 & 2.8 & 2.3 & 2.1\\
	& $\times 10^{-15} / 100 \mathrm{~s}$ & 1.3 & 0.7 & 1.1 & 1.5 & 1.0 \\
      \hline
		\end{tabular}}
\end{table}

\section{Impact of frequency tampering}
Frequency reference synchronization is important in timekeeping, metrology, astronomical observation, etc. If the recovered frequency is covertly altered, severe hazards can occur. In the following, we present three frequency tampering scenarios, where the function of these applications is still normal, but the results provided are highly biased.

\begin{figure}[h]
	\centering
	\includegraphics[width=3.4in]{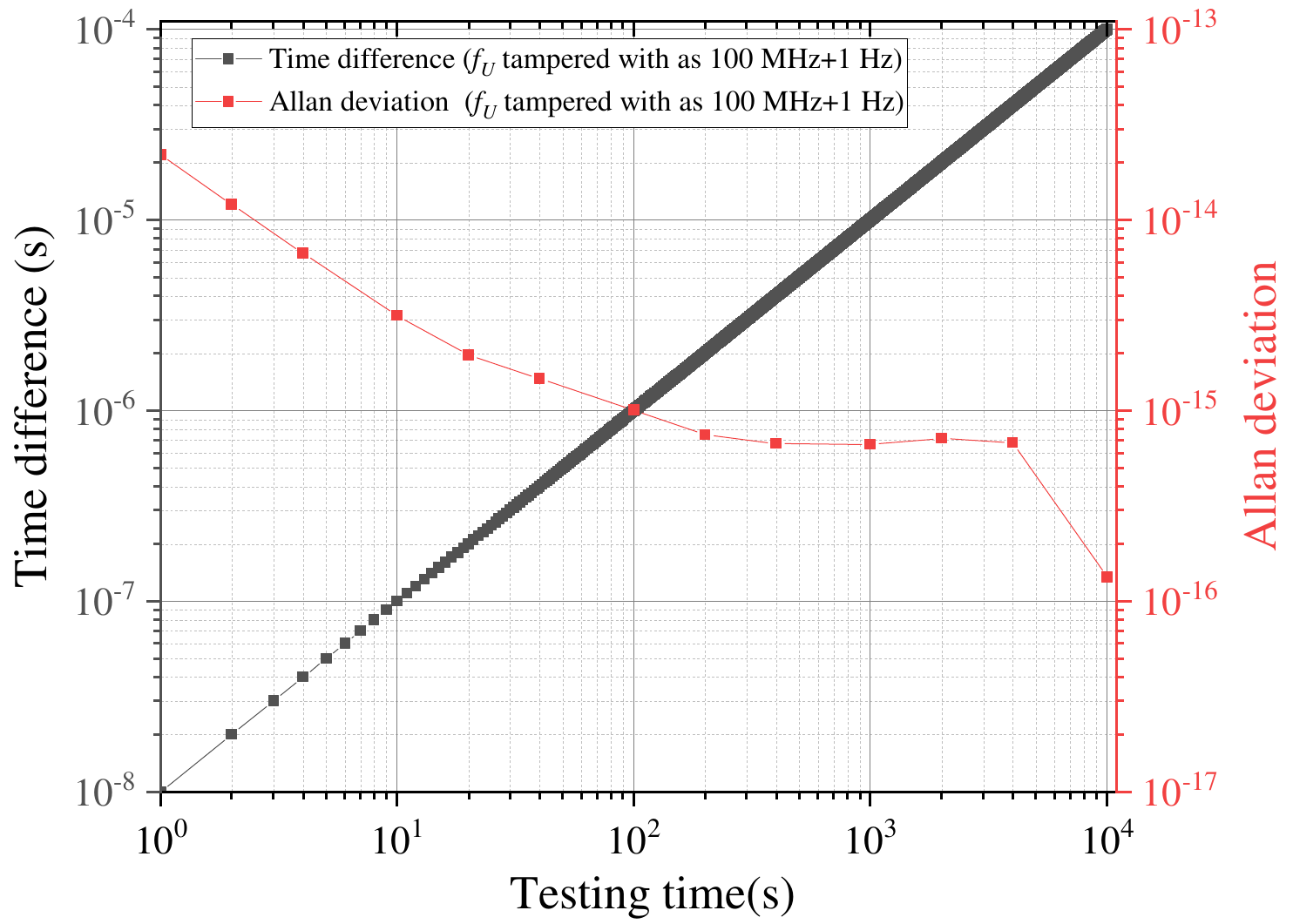}		
	\caption{Impact of frequency tampering on the timekeeping system. The recovery frequency signal to be used as the reference for the user-side timekeeping system is altered from 100 MHz to 100 MHz+1 Hz.}
	\label{fig7}
\end{figure}

For the case of the timekeeping network, the frequency tampering directly affects the length of second \cite{Riehle2017}, due to the relationship of $\frac{\Delta t}{t}=\frac{\Delta f}{f}$. Considering a scenario of two timekeeping systems at the server and user of Fig. \ref{fig3}, if the recovered frequency at the user side is altered from 100 MHz to 100 MHz+1 Hz, the time difference between 2 timekeeping systems increases 0.1 ms every $10^4$ s, as shown in Fig. \ref{fig7}. From the viewpoint of the user side, its timekeeping function is normal because the timing output is stable and reproducible, which is found in the Allan deviation plot of Fig. \ref{fig7}.  From the viewpoint of the whole timekeeping network, the cumulated time difference destroys the synchronization of the timekeeping network.

\begin{figure}[!h]
	\centering
	\includegraphics[width=3.1in]{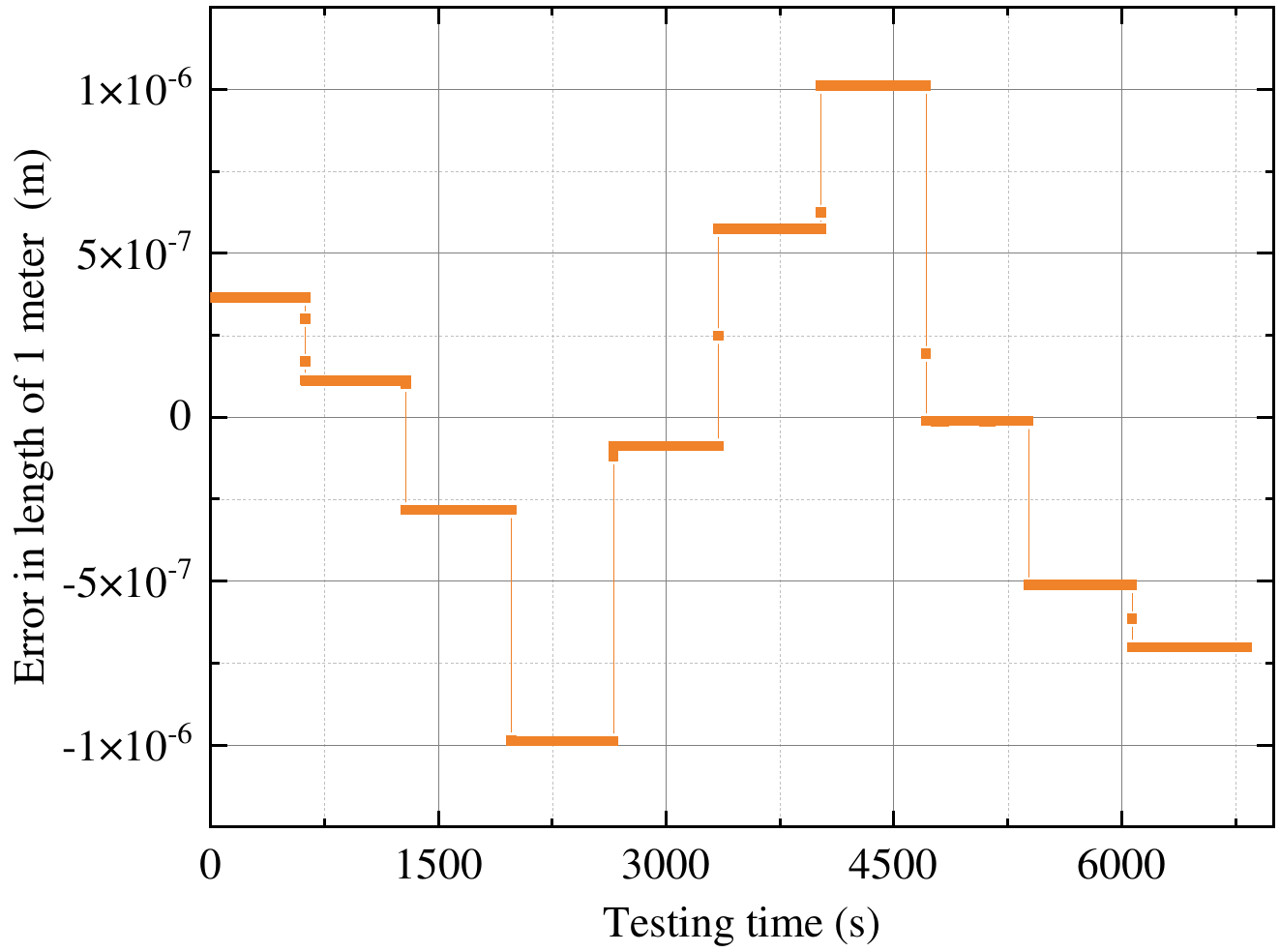}
	\caption{Simulation result of the corresponding error of the recovered SI unit-meter using the tampered frequency signal shown in Fig. \ref{fig6} as the reference for the unit of length tracing.}
	\label{fig8}
\end{figure}

For the new definition of the International System of Units (SI) \cite{stock2019}, all SI units (except for mole and second itself) can be traced to corresponding fundamental constants with the aid of second, which is the most precise SI unit. The unit of length is defined by using the fixed numerical value of the speed of light in vacuum c to be 299792458 expressed in m/s. If the recovered frequency in Fig. \ref{fig6} is used as the reference of the meter recovery system, the tampering action on frequency transfers to the tampering action on the length. As shown in Fig. \ref{fig8}, the corresponding error of the recovered SI unit-meter is altered with within the range of $\pm 1 \times 10^{-6}$ m. In this case, the function of the meter recovery system remains normal, while the induced by frequency tampering causes a large error.

Another scenario is radio astronomy. In radio astronomy, multiple radio telescopes can be combined using radio interferometry \cite{shuang2023,Holler2012} and aperture synthesis techniques \cite{Le1999} to improve measurement accuracy. To reduce the coherent loss of the observation system, interference measurement arrays, such as the Square Kilometre Array (SKA), have a high requirement for frequency synchronization \cite{Alachkar2018}. If there is a fixed frequency difference $\Delta f$ between the two observation 

\begin{figure}[h]
	\centering
	\includegraphics[width=3in]{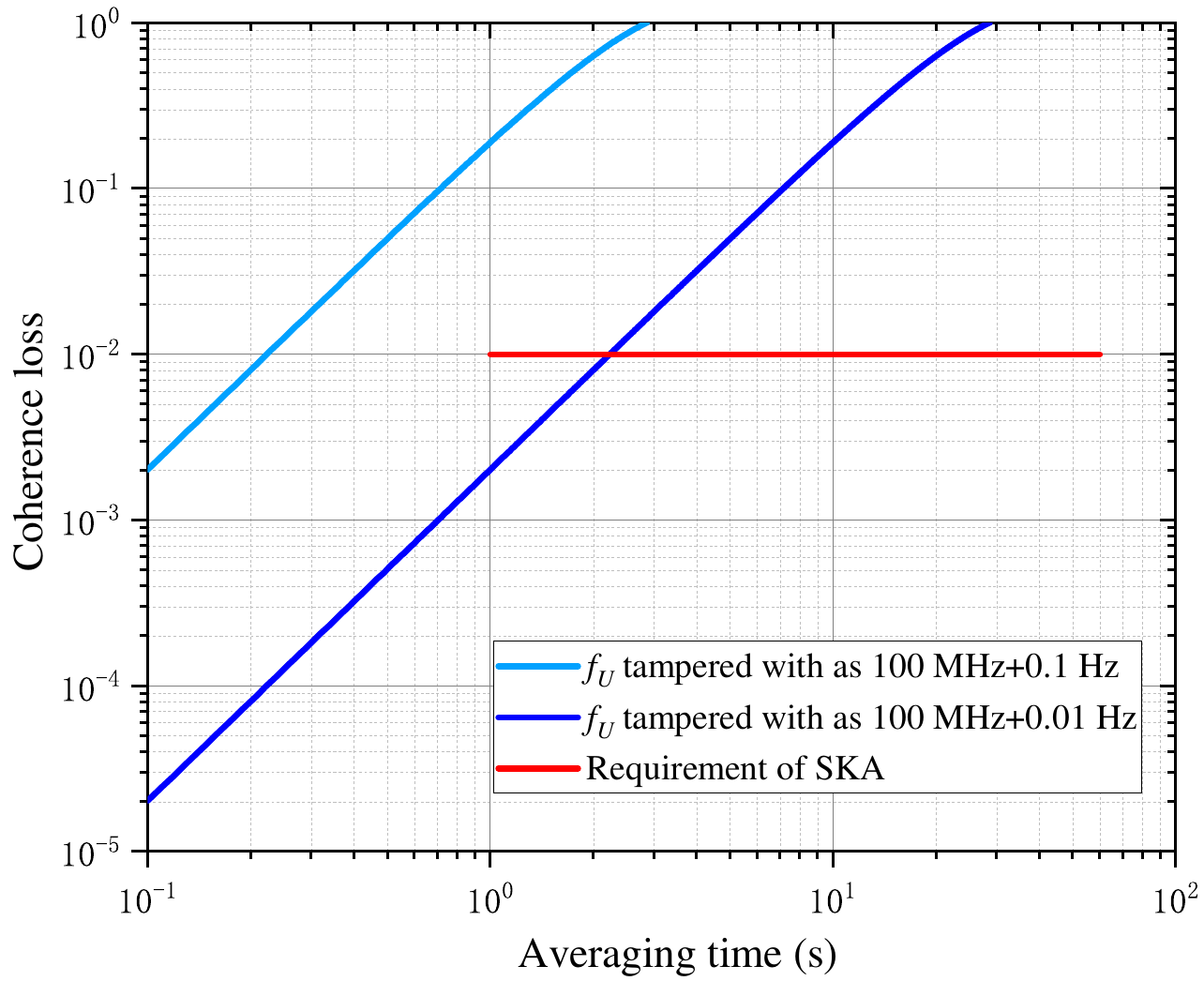}
	\caption{Coherence loss of the radio telescope array under different frequency tampering cases, which are compared with the requirements of the Square Kilometre Array (SKA).}
	\label{fig9}
\end{figure}

\noindent stations caused by the FTM, the coherent loss $Lc(T)$ between the two stations follows the relationship: $Lc(T)=1-|\operatorname{sinc}(\pi\Delta ft)|$. For example, when the observation frequency is 350 MHz, the calculation results of the coherent loss under different frequency tampering scenarios are shown in Fig. \ref{fig9}. To compare the impact of reference frequency tampering directly, we use the coherent loss requirement of SKA as a comparison. SKA requires that the coherent loss caused by the frequency transmission system does not exceed $1\% $ within an integration time from 1 s to 1 min. The coherent loss caused by 0.01 Hz frequency tampering causes the SKA observation to be disabled at 2.5 s, and the observation is completely unfeasible at 0.1 Hz frequency tampering.

\section{Discussion}
The FbFS network is one of the most important infrastructures. However, the frequency-tampering scenarios expose its potential risks. To solve its security issue, countermeasures of this risk posed by the FTM need to be discussed.

In optical networks, signal encryption \cite{Venu2022}and link integrity monitoring \cite{Bao2012,Williams2019} are commonly used security measures at the information and hardware layers, respectively. Unfortunately, neither of these methods cannot be used as security countermeasures for the disseminated frequency reference. First, signal encryption techniques have a negative impact on the relative stability of the disseminated frequency signal and introducing additional phase noise, which defeats the purpose of high-quality transmission of frequency references.
Second, link integrity monitoring methods, represented by optical time domain reflectometry (OTDR) \cite{Lu2010,Peng2014,Peng2023}, can be circumvented by the FTM. DWDMs and controlled optical switches can ensure that the core structure of the FTM is not detected during OTDR detection. All these circumstances point to the need to identify new security countermeasures for FbFS.

There are two possible security countermeasures. One countermeasure is link patrolling combined with fiber vibration detection. Relying on fiber vibration sensing technology, such as distributed acoustic sensing (DAS) \cite{li2021,van2023,Lindsey2019,Wang2022}, it is possible to identify intrusions and quickly locate them. Every link interruption event needs to be taken seriously. Inspectors should inspect risky sections and ensure that no special devices, such as FTMs, are embedded in the link. Another countermeasure is to build a more robust FbFS network. Multiple links need to be established between a single server and a single user. Preferably, the user can receive frequency references disseminated by different servers. If the user can receive multiple frequency references, they can be compared with each other to ensure that problems with a particular frequency reference can be detected in time. 

\section{Conclusion}
On a 200 km fiber link, we demonstrate that the synchronized frequency reference can be altered using a frequency lens-enabled FTM. While bypassing the locking status monitoring, the 100 MHz frequency reference synchronized at the user side can be changed arbitrarily and stealthily within the range of 100 MHz±100 Hz. Moreover, the relative stability of the recovered frequency reference can be maintained at the same normal level when tampering occurs. To more intuitively show the consequences of using a tampered frequency reference, we constructed three application scenarios that rely on disseminated frequency reference. In these three scenarios, the application systems could still run stably, but the generated results deviated dramatically. To solve this tampering risk, we propose two possible countermeasures. We anticipate that this article will raise awareness of the risks faced by FbFS networks and improve risk prevention measures.

\appendix

\setcounter{equation}{0}
\renewcommand\theequation{A\arabic{equation}} 
\section*{Appendix A: Details of the fiber-based radio frequency synchronization system and derivation process of Eq. (10)}\label{appendix1}

To show the details of the fiber-based radio frequency synchronization system as shown in Fig. \ref{fig3}(a), we start the explanation from PLL1 and PLL2. For convenience, the amplitude terms of signals in Appendix A are also ignored. A 100 MHz voltage-controlled crystal oscillator (VCXO) at the server site generates the signal that can be expressed as $V_{SX}=V_{VCXO-S}=\cos(2\pi f_{SX} t+\phi_{SX})$. Then, $V_{SX}$ is converted to 2.1 GHz by a phase-locked dielectric resonant oscillator (PDRO), i.e., $V_{PDRO-S}=\cos(2\pi \times 21f_{SX} t+21\phi_{SX})$, which is modulated to an optical carrier (C38), and sent to the user site via the fiber link.

In the following, we first analyze the normal operation of the system when the FTM does not join the link. At the user site, the optical signal transmitted along the fiber is received and demodulated. The 2.1 GHz RF signal can be obtained at U1 point of Fig. \ref{fig3}(a), i.e., $V_{U1}=\cos(2\pi \times 21f_{SX} t+21\phi_{SX}+\phi_n)$, where $\phi_n$ is the phase noise caused by the whole link. Another 100 MHz VCXO-U located at the user site generates a signal that can be expressed as $V_{UX}=V_{VCXO-U}=\cos(2\pi  f_{UX} t+\phi_{UX})$. Similarly, $V_{UX}$ is also multiplied to 2.1GHz by the PDRO-U, i.e., $V_{PDRO-U}=\cos(2\pi \times 21f_{UX} t+21\phi_{UX})$. $V_{PDRO-U}$ is mixed with $V_{U1}$ to generate the error signal, i.e., $V_{error-U}=\cos[2\pi (21f_{SX}-21f_{UX} )t+21(\phi_{SX}-\phi_{UX})+\phi_n]$. Then, PLL2 can be realized with a servo controller. As a consequence,
\begin{equation}    
\begin{aligned}
	21f_{SX}-21f_{UX}=0,\label{eqA1}\\
 \end{aligned}
\end{equation}
\begin{equation}    
\begin{aligned}
	21\phi_{SX}+\phi_n-21\phi_{UX}=0. \qquad\ \label{eqA2}
\end{aligned}
\end{equation}
For the backward route, $V_{PDRO-U}$ is modulated on another optical carrier (C39), and sent to the server site. After demodulation and amplification at the server site, a 2.1 GHz RF signal can be obtained at point M1. In addition, the phase noise introduced by the same fiber link for a short time can also be considered as $\phi_n$ too. Considering the above, the 2.1 GHz RF signal at M1 point can be expressed as $V_{M1}=\cos(2\pi \times 21 f_{UX} t+21\phi_{UX}+\phi_n)$.

In the left part of Fig. \ref{fig3}(a), the reference frequency signal of 100 MHz, i.e., $V_{ref}=\cos(2\pi f_{ref} t+\phi_{ref})$, is split and sent to two PDROs to generate the RF signals of 2.2 GHz and 2.0 GHz, denoted as $V_{ref1}=V_{PDRO-1}=\cos(2\pi \times 22f_{ref} t+22\phi_{ref})$ and $V_{ref2}=V_{PDRO-2}=\cos(2\pi \times 20f_{ref} t+20\phi_{ref})$, respectively. Subsequently, $V_{ref1}$ is mixed with $V_{M1}$ to generate a 100 MHz signal, i.e., $\cos[2\pi (22f_{ref}-21 f_{UX})t+(22\phi_{ref}-21\phi_{UX})-\phi_n]$. Similarly, $V_{ref2}$ is mixed with $V_{M2}=V_{PDRO-S}$ to generate another 100 MHz signal, i.e., $\cos[2\pi (21f_{SX}-20f_{ref})t+21\phi_{SX}-20\phi_{ref}]$. Finally, these two 100 MHz signals are mixed to obtain the error signal $V_{error-S}$, i.e., $V_{error-S}=\cos[2\pi (42f_{ref}-21f_{SX}-21f_{UX})t+(42\phi_{ref}-21\phi_{SX}-21\phi_{UX}-\phi_n)]$. PLL1 can be realized with a servo controller. As a consequence, 
\begin{equation}    
\begin{split}
42f_{ref}-21f_{SX}-21f_{UX}=0,\label{eqA3}\\    
\end{split}
\end{equation}
\begin{equation}    
\begin{split}
42\phi_{ref}-21\phi_{SX}-21\phi_{UX}-\phi_n=0.\label{eqA4}
\end{split}
\end{equation}
Considering Eq. (\ref{eqA1}), Eq. (\ref{eqA2}), Eq. (\ref{eqA3}), and Eq. (\ref{eqA4}), the output signal $V_{Output}$ at the user site can be obtained as
\begin{equation}    
\begin{aligned}
	V_{Output}&=V_{UX}=V_{VCXO-U}\\
	&=\cos(2\pi f_{UX} t+\phi_U )\\
	& =\cos(2\pi f_{ref} t+\phi_{ref} ),\label{eqA5}
 \end{aligned}
\end{equation}
and frequency synchronization between the server site and the user site is realized.

In the following we consider the case after inserting the FTM and provide the derivation process of Eq. \eqref{eq10}. 
Before entering the FTM, the RF signal carried by the optical signal (C38) is $\cos(2\pi \times 21f_{SX} t+21\phi_{SX}+\phi_{n'1})$, where $\phi_{n'1}$ is the noise introduced by part of the fiber link between the FTM and the server site. Referring to Eq. \eqref{eq5} and Eq. \eqref{eq6}, after passing the FTM, the RF signal carried by the optical signal (C38) is $\cos[2\pi (k_1 f_{S0}+21f_{SX})t+k_1 \phi_{S0}+21\phi_{SX}+\phi_{n'1}]$. As a consequence, $V_{U1}$ changes, i.e., $V_{U1}=\cos[2\pi (k_1 f_{S0}+21f_{SX} )t+k_1 \phi_{S0}+21\phi_{SX}+\phi_{n'1}+\phi_{n'2}]$, where $\phi_{n'2}$ is the noise introduced by another part of the fiber link. Due to PLL2, $V_{UX}$ changes accordingly, i.e., 
\begin{equation}    
\begin{aligned}
	f_{UX} &=\frac{k_1 f_{S0}}{21} + f_{SX},\qquad\qquad\qquad\ \label{eqB1}\\
 \end{aligned}
\end{equation}
\begin{equation}    
\begin{aligned}
	\phi_{UX} &=\frac{k_1 \phi_{S0}}{21}+\phi_{SX}+\frac{\phi_{n'1}+\phi_{n'2}}{21}.\label{eqB2}
\end{aligned}
\end{equation}
Therefore, the RF signal modulated to laser-2 changes to $V_{PDRO-U}=\cos[2\pi (k_1 f_{S0}+21f_{SX} )t+k_1 \phi_{S0}+21\phi_{SX}+\phi_{n'1}+\phi_{n'2} ]$. In the backward direction, when entering the FTM, the RF signal carried by the optical signal (C39) is $\cos[2\pi (k_1 f_{S0}+21f_{SX} )t+k_1 \phi_{S0}+21\phi_{SX}+\phi_{n'1}+2\phi_{n'2}]$. Referring to Eq. \eqref{eq8} and Eq. \eqref{eq9}, the RF signal modulated on the optical carrier changes back to $\cos[2\pi \times 21f_{SX} t+(21\phi_{SX}+\phi_{n'1}+2\phi_{n'2})]$. Although the FTM tampers with the disseminated frequency reference in the fiber link, the frequency of the RF signal demodulated at the server site remains unchanged. After going through the same process described by Eq. \eqref{eqA3} and Eq. \eqref{eqA4}, what can be obtained is
\begin{equation}    
\begin{aligned}
	f_{SX} &=f_{ref},\qquad\qquad\qquad\ \label{eqB3}\\
 \end{aligned}
\end{equation}
\begin{equation}    
\begin{aligned}
    \phi_{SX} &=\phi_{ref}-\frac{\phi_{n'1}+\phi_{n'2}}{21}. \label{eqB4}
\end{aligned}
\end{equation}
As a result, 
\begin{equation}    
\begin{aligned}
	V_{Output}&=\cos(2\pi f_{UX} t+\phi_{UX} )\\
	&=\cos\left[ 2\pi \left( \frac{k_1 f_{S0}}{21}+f_{ref} \right)t+\frac{k_1 \phi_{S0}}{21}+\phi_{ref} \right]. \label{eqB5}
\end{aligned}
\end{equation}
In the experimental setup, $f_{S0}=10$ MHz and $f_{ref}=100$ MHz, so that 
\begin{eqnarray}
	f_{UX}=\left( 100+ \frac{10k_1}{21} \right) \textnormal{MHz}.\label{eqB6}
\end{eqnarray}


\bibliographystyle{IEEEtran}
\bibliography{IEEEabrv,refer}

\begin{thebibliography}{10}
\providecommand{\url}[1]{#1}
\csname url@samestyle\endcsname
\providecommand{\newblock}{\relax}
\providecommand{\bibinfo}[2]{#2}
\providecommand{\BIBentrySTDinterwordspacing}{\spaceskip=0pt\relax}
\providecommand{\BIBentryALTinterwordstretchfactor}{4}
\providecommand{\BIBentryALTinterwordspacing}{\spaceskip=\fontdimen2\font plus
\BIBentryALTinterwordstretchfactor\fontdimen3\font minus \fontdimen4\font\relax}
\providecommand{\BIBforeignlanguage}[2]{{%
\expandafter\ifx\csname l@#1\endcsname\relax
\typeout{** WARNING: IEEEtran.bst: No hyphenation pattern has been}%
\typeout{** loaded for the language `#1'. Using the pattern for}%
\typeout{** the default language instead.}%
\else
\language=\csname l@#1\endcsname
\fi
#2}}
\providecommand{\BIBdecl}{\relax}
\BIBdecl

\bibitem{Ye2003}
J.~Ye, J.~Peng, R.~J. Jones, K.~W. Holman, J.~L. Hall, D.~J. Jones, S.~A. Diddams, J.~Kitching, S.~Bize, J.~C. Bergquist, L.~W. Hollberg, L.~Robertsson, and L.~S. Ma, ``{Delivery of high-stability optical and microwave frequency standards over an optical fiber network},'' \emph{J. Opt. Soc. Am. B}, vol.~20, pp. 1459--1467, 2003.

\bibitem{Williams2008}
P.~A. Williams, W.~C. Swann, and N.~R. Newbury, ``{High-stability transfer of an optical frequency over long fiber-optic links},'' \emph{J. Opt. Soc. Am. B}, vol.~25, pp. 1284--1293, Aug. 2008.

\bibitem{Masaki2008}
M.~Amemiya, M.~Imae, Y.~Fujii, T.~Suzuyama, and S.-i. Ohshima, ``Simple time and frequency dissemination method using optical fiber network,'' \emph{IEEE Trans. Instrum. Meas.}, vol.~57, no.~5, pp. 878--883, 2008.

\bibitem{Wang2012}
B.~Wang, C.~Gao, W.~L. Chen, J.~Miao, X.~Zhu, Y.~Bai, J.~W. Zhang, Y.~Y. Feng, T.~C. Li, and L.~J. Wang, ``{Precise and continuous time and frequency synchronisation at the $5\times 10^{-19}$ Accuracy Level},'' \emph{Sci. Rep.}, vol.~2, p. 556, 2012.

\bibitem{Bai2013}
Y.~Bai, B.~Wang, X.~Zhu, C.~Gao, J.~Miao, and L.~J. Wang, ``{Fiber-based multiple-access optical frequency dissemination},'' \emph{Opt. Lett.}, vol.~38, pp. 3333--3335, 2013.

\bibitem{Lisdat2016}
C.~Lisdat, G.~Grosche, N.~Quintin, C.~Shi, S.~Raupach, C.~Grebing, D.~Nicolodi, F.~Stefani, A.~Al-Masoudi, S.~D{\"o}rscher \emph{et~al.}, ``A clock network for geodesy and fundamental science,'' \emph{Nat. Commun.}, vol.~7, no.~1, p. 12443, 2016.

\bibitem{Rizzi2016}
M.~Rizzi, S.~Rinaldi, P.~Ferrari, A.~Flammini, D.~Fontanelli, and D.~Macii, ``Enhancing accuracy and robustness of frequency transfer using synchronous ethernet and multiple network paths,'' \emph{IEEE Trans. Instrum. Meas.}, vol.~65, no.~8, pp. 1926--1936, 2016.

\bibitem{Riehle2017}
F.~Riehle, ``Optical clock networks,'' \emph{Nat. Photonics}, vol.~11, no.~1, pp. 25--31, 2017.

\bibitem{Wang2020JL}
J.~Wang, C.~Yue, Y.~Xi, Y.~Sun, N.~Cheng, F.~Yang, M.~Jiang, J.~Sun, Y.~Gui, and H.~Cai, ``Fiber-optic joint time and frequency transfer with the same wavelength,'' \emph{Opt. Lett.}, vol.~45, no.~1, pp. 208--211, 2020.

\bibitem{Liu2022Bo}
B.~Liu, X.~Guo, W.~Kong, T.~Liu, R.~Dong, and S.~Zhang, ``Stabilized time transfer via a 1000-km optical fiber link using high-precision delay compensation system,'' \emph{Photonics}, vol.~9, no.~8, 2022.

\bibitem{Schioppo2022}
M.~Schioppo, J.~Kronjaeger, A.~Silva, R.~Ilieva, J.~Paterson, C.~Baynham, W.~Bowden, I.~Hill, R.~Hobson, A.~Vianello \emph{et~al.}, ``Comparing ultrastable lasers at 7$\times 10^{-17}$ fractional frequency instability through a 2220 km optical fibre network,'' \emph{Nat. Commun.}, vol.~13, no.~1, p. 212, 2022.

\bibitem{johnson2020}
S.~Johnson and O.~A. Dobre, ``Time and carrier frequency synchronization for coherent optical communication: Implementation considerations, measurements, and analysis,'' \emph{IEEE Trans. Instrum. Meas.}, vol.~69, no.~8, pp. 5810--5820, 2020.

\bibitem{Lewis2021}
S.~Lewis and M.~Inggs, ``Synchronization of coherent netted radar using white rabbit compared with one-way multichannel gpsdos,'' \emph{IEEE Trans. Aerosp. Electron. Syst.}, vol.~57, no.~3, pp. 1413--1422, 2021.

\bibitem{He2018}
Y.~He, K.~G.~H. Baldwin, B.~J. Orr, R.~Bruce~Warrington, M.~J. Wouters, A.~N. Luiten, P.~Mirtschin, T.~Tzioumis, C.~Phillips, J.~Stevens, B.~Lennon, S.~Munting, G.~Aben, T.~Newlands, and T.~Rayner, ``{Long-distance telecom-fiber transfer of a radio-frequency reference for radio astronomy},'' \emph{Optica}, vol.~5, pp. 138--146, 2018.

\bibitem{Chen2021}
Y.~Chen, B.~Wang, L.~Wang, K.~Grainge, R.~Oberland, R.~Whitaker, and A.~Wilkinson, ``Integrated dissemination system of frequency, time and data for radio astronomy,'' \emph{IEEE Photon. J.}, vol.~13, no.~1, pp. 1--7, 2021.

\bibitem{Lukasz2020}
L.~Sliwczynski, P.~Krehlik, H.~Imlau, H.~Ender, H.~Schnatz, D.~Piester, and A.~Bauch, ``Fiber-based utc dissemination supporting 5g telecommunications networks,'' \emph{IEEE Commun. Mag.}, vol.~58, no.~4, pp. 67--73, 2020.

\bibitem{Zhou2023}
Z.~Zhou, J.~Wei, Y.~Luo, K.~A. Clark, E.~Sillekens, C.~Deakin, R.~Sohanpal, R.~Slavik, and Z.~Liu, ``Communications with guaranteed bandwidth and low latency using frequency-referenced multiplexing,'' \emph{Nat. Electron.}, vol.~6, no.~9, pp. 694--702, 2023.

\bibitem{Marra2018}
G.~Marra, C.~Clivati, R.~Luckett, A.~Tampellini, J.~Kronjager, L.~Wright, A.~Mura, F.~Levi, S.~Robinson, A.~Xuereb, B.~Baptie, and D.~Calonico, ``{Ultrastable laser interferometry for earthquake detection with terrestrial and submarine cables},'' \emph{Science}, vol. 361, p. 486, 2018.

\bibitem{Grotti2018}
J.~Grotti, S.~Koller, S.~Vogt, S.~H{\"a}fner, U.~Sterr, C.~Lisdat, H.~Denker, C.~Voigt, L.~Timmen, A.~Rolland \emph{et~al.}, ``Geodesy and metrology with a transportable optical clock,'' \emph{Nat. Phys.}, vol.~14, no.~5, pp. 437--441, 2018.

\bibitem{Clark2020}
K.~A. Clark, D.~Cletheroe, T.~Gerard, I.~Haller, K.~Jozwik, K.~Shi, B.~Thomsen, H.~Williams, G.~Zervas, H.~Ballani, P.~Bayvel, P.~Costa, and Z.~Liu, ``Synchronous subnanosecond clock and data recovery for optically switched data centres using clock phase caching,'' \emph{Nat. Electron.}, vol.~3, no.~7, pp. 426--433, 2020.

\bibitem{narula2018}
L.~Narula and T.~E. Humphreys, ``Requirements for secure clock synchronization,'' \emph{IEEE Journal of Selected Topics in Signal Processing}, vol.~12, no.~4, pp. 749--762, 2018.

\bibitem{lee2019}
J.~Lee, L.~Shen, A.~Cer{\`e}, J.~Troupe, A.~Lamas-Linares, and C.~Kurtsiefer, ``Asymmetric delay attack on an entanglement-based bidirectional clock synchronization protocol,'' \emph{Appl. Phys. Lett.}, vol. 115, no.~14, 2019.

\bibitem{Li2023}
Y.~Li, J.~Hu, Y.~Pan, W.~Huang, L.~Ma, J.~Yang, S.~Zhang, Y.~Luo, C.~Zhou, C.~Zhang, H.~Wang, Y.~Shao, Y.~Zhang, X.~Chen, Z.~Chen, S.~Yu, H.~Guo, and B.~Xu, ``Secure two-way fiber-optic time transfer against sub-ns asymmetric delay attack with clock model-based detection and mitigation scheme,'' \emph{IEEE Trans. Instrum. Meas.}, vol.~72, 2023.

\bibitem{Xu2023}
X.~Xu, Y.~Zhang, Y.~Bian, J.~Hu, J.~Dou, Y.~Li, B.~Xu, S.~Yu, and H.~Guo, ``Controllable asymmetric attack against practical round-trip fiber time synchronization systems,'' \emph{IEEE Photon. Technol. Lett.}, pp. 1263--6, 2023.

\bibitem{Zhang2021}
C.~Zhang, Y.~Li, X.~Chen, Y.~Zhang, L.~Fu, Y.~Gong, H.~Wang, W.~Huang, and B.~Xu, ``Controllable asymmetry attack on two-way fiber time synchronization system,'' \emph{IEEE Photon. J.}, vol.~13, no.~6, 2021.

\bibitem{Liu2022}
Z.~Liu, Y.~Bian, Y.~Zhang, B.~Xu, Y.~Li, and S.~Yu, ``Asymmetric channel attack against practical round-trip fiber time synchronization system,'' in \emph{2022 Joint Conference of the Europenan Frequency and Time Forum and IEEE International Frequency Control Symposium (EFTF/IFCS)}, ser. Joint European Frequency and Time Forum and International Frequency Control Symposium, 2022.

\bibitem{Chen2022}
Y.~Chen, H.~Dai, H.~Si, F.~Wang, B.~Wang, and L.~Wang, ``Long-haul high precision frequency dissemination based on dispersion correction,'' \emph{IEEE Trans. Instrum. Meas.}, vol.~71, pp. 1--7, 2022.

\bibitem{stock2019}
M.~Stock, R.~Davis, E.~de~Mirand{\'e}s, and M.~J. Milton, ``The revision of the si—the result of three decades of progress in metrology,'' \emph{Metrologia}, vol.~56, no.~2, p. 022001, 2019.

\bibitem{shuang2023}
W.-H. Shang-Guan, R.-B. Zhao, J.-Q. Wang, Z.-C. Wang, Q.-H. Liu, X.-Y. Hong, G.-L. Wang, W.-M. Zheng, X.-Z. Zhang, T.~Shuai, Z.~Yan, Y.-D. Huang, X.-J. Lu, L.-F. Yu, Y.-B. Jiang, C.~Zhang, M.-L. Ma, W.-Y. Zhong, R.-J. Zhu, W.-B. Wang, J.~Zhang, B.~Xia, and C.-Y. Zhang, ``Lunar orbit vlbi experiment ground validation system,'' \emph{IEEE Trans. Instrum. Meas.}, vol.~72, pp. 1--14, 2023.

\bibitem{Holler2012}
C.~M. Holler, M.~E. Jones, A.~C. Taylor, A.~I. Harris, and S.~A. Maas, ``A 2–20-ghz analog lag correlator for radio interferometry,'' \emph{IEEE Trans. Instrum. Meas.}, vol.~61, no.~8, pp. 2253--2261, 2012.

\bibitem{Le1999}
D.~Le~Vine, ``Synthetic aperture radiometer systems,'' \emph{IEEE Trans Microw Theory Tech}, vol.~47, no.~12, pp. 2228--2236, 1999.

\bibitem{Alachkar2018}
B.~Alachkar, A.~Wilkinson, and K.~Grainge, ``Frequency reference stability and coherence loss in radio astronomy interferometers application to the ska,'' \emph{J. Astron. Instrum.}, vol.~7, no.~01, p. 1850001, 2018.

\bibitem{Venu2022}
D.~Venu, A.~Mayuri, S.~Neelakandan, G.~Murthy, N.~Arulkumar, and N.~Shelke, ``An efficient low complexity compression based optimal homomorphic encryption for secure fiber optic communication,'' \emph{Optik}, vol. 252, p. 168545, 2022.

\bibitem{Bao2012}
X.~Bao and L.~Chen, ``Recent progress in distributed fiber optic sensors,'' \emph{SENSORS}, vol.~12, no.~7, pp. 8601--8639, 2012.

\bibitem{Williams2019}
E.~F. Williams, M.~R. Fern{\'a}ndez-Ruiz, R.~Magalhaes, R.~Vanthillo, Z.~Zhan, M.~Gonz{\'a}lez-Herr{\'a}ez, and H.~F. Martins, ``Distributed sensing of microseisms and teleseisms with submarine dark fibers,'' \emph{Nat. Commun.}, vol.~10, no.~1, p. 5778, 2019.

\bibitem{Lu2010}
Y.~Lu, T.~Zhu, L.~Chen, and X.~Bao, ``Distributed vibration sensor based on coherent detection of phase-otdr,'' \emph{J. Light. Technol.}, vol.~28, no.~22, pp. 3243--3249, 2010.

\bibitem{Peng2014}
F.~Peng, H.~Wu, X.~H. Jia, Y.~J. Rao, Z.~N. Wang, and Z.~P. Peng, ``{Ultra-long high-sensitivity phi-OTDR for high spatial resolution intrusion Detection of pipelines},'' \emph{Opt. Express}, vol.~22, pp. 13\,804--13\,810, 2014.

\bibitem{Peng2023}
F.~Peng, X.~Zheng, and Q.~Miao, ``Large dynamic range and anti-fading phase-sensitive otdr using 2-d phase unwrapping via neural network,'' \emph{IEEE Trans. Instrum. Meas.}, vol.~72, pp. 1--8, 2023.

\bibitem{li2021}
T.~Li, C.~Fan, H.~Li, T.~He, W.~Qiao, Z.~Shi, Z.~Yan, C.~Liu, D.~Liu, and Q.~Sun, ``Nonintrusive distributed flow rate sensing system based on flow-induced vibrations detection,'' \emph{IEEE Trans. Instrum. Meas.}, vol.~70, pp. 1--8, 2021.

\bibitem{van2023}
M.~van~den Ende, I.~Lior, J.-P. Ampuero, A.~Sladen, A.~Ferrari, and C.~Richard, ``A self-supervised deep learning approach for blind denoising and waveform coherence enhancement in distributed acoustic sensing data,'' \emph{IEEE Trans. Instrum. Meas.}, vol.~34, no.~7, pp. 3371--3384, 2023.

\bibitem{Lindsey2019}
N.~J. Lindsey, T.~C. Dawe, and J.~B. Ajo-Franklin, ``{Illuminating seafloor faults and ocean dynamics with dark fiber distributed acoustic sensing},'' \emph{Science}, vol. 366, p. 1103, 2019.

\bibitem{Wang2022}
G.~Wang, Z.~W. Pang, B.~H. Zhang, F.~M. Wang, Y.~F. Chen, H.~F. Dai, B.~Wang, and L.~J. Wang, ``{Time shifting deviation method enhanced laser interferometry: ultrahigh precision localizing of traffic vibration using an urban fiber link},'' \emph{Photonics Res.}, vol.~10, pp. 1839--1839, 2022.

\end{thebibliography}

\end{document}